\documentstyle[12pt,epsf]{article}
\renewcommand{\bar}[1]{\overline{#1}}

\begin{document}

\begin{flushright}
USM-TH-112\\
CPT-2001/P.4238
\end{flushright}
\bigskip\bigskip

\centerline{\Large \bf Quark Distributions of Octet Baryons from SU(3) Symmetry}

 \vspace{22pt} \centerline{\bf
Bo-Qiang Ma\footnote{e-mail: mabq@phy.pku.edu.cn}$^{a}$,
Ivan Schmidt\footnote{e-mail: ischmidt@fis.utfsm.cl}$^{b}$,
Jacques Soffer\footnote{e-mail: Jacques.Soffer@cpt.univ-mrs.fr}$^{c}$,
and Jian-Jun Yang\footnote{e-mail: jjyang@fis.utfsm.cl}$^{b,d}$}

\vspace{8pt}

{\centerline {$^{a}$Department of Physics, Peking University,
Beijing 100871, China\footnote{Mailing address},}}

{\centerline {CCAST (World Laboratory),
P.O.~Box 8730, Beijing 100080, China,}}

{\centerline {and Institute of Theoretical Physics, Academia
Sinica, Beijing 100080, China}

{\centerline {$^{b}$Departamento de F\'\i sica, Universidad
T\'ecnica Federico Santa Mar\'\i a,}}

{\centerline {Casilla 110-V, 
Valpara\'\i so, Chile}

{\centerline {$^{c}$Centre de Physique Th$\acute{\rm{e}}$orique,
CNRS, Luminy Case 907,}}
{\centerline { F-13288 Marseille Cedex 9, France}}

{\centerline {$^{d}$Department of Physics, Nanjing Normal
University,}}

{\centerline {Nanjing 210097, China}}

\vspace{10pt}
\begin{center} {\large \bf Abstract}
\end{center}

SU(3) symmetry relations between the octet baryons are introduced
in order to connect both the unpolarized and polarized quark distributions of the  
octet baryons with those of the nucleon.
Two different parametrizations of the nucleon quark distributions are used. 
A new scenario of quark flavor and spin structure of the $\Lambda$ is
found and compared with two other models: a perturbative QCD based analysis and
a quark diquark model. The $u$ and $d$ quarks inside the $\Lambda$ are predicted
to be positively polarized at large Bjorken variable $x$ in the new scenario.
By using an approximate relation connecting the quark fragmentation functions
with the quark distributions, the hadron polarizations of the octet baryons in
$e^+e^-$-annihilation,
polarized charged lepton deep inelastic scattering (DIS) processes,
and neutrino (antineutrino) DIS processes are predicted.
The predictions for $\Lambda$ polarizations in several processes are compatible
with the available data at large fragmentation momentum fraction $z$,
and support the prediction
of positively polarized $u$ and $d$ quarks inside the $\Lambda$ at large $x$.
Predictions for Drell-Yan processes from  
$\Sigma^{\pm}$ and $\Xi^-$ beams on an isoscalar target are also given and discussed.

\vfill \centerline{PACS numbers: 14.20.Jn, 13.65.i, 13.87.Fh, 13.88.+e}

\vfill

\newpage

\section{Introduction}

Quark distributions are important basic physical quantities containing the
information of the underlying structure of the nucleon, and they can be
precisely measured by combining various deep inelastic scattering (DIS)
\cite{DISprocess} and Drell-Yan processes \cite{DYprocess}.
After more than three decades of experimental
studies, our knowledge of the quark distributions for the nucleon
is more or less clear concerning
the bulk features of momentum, flavor and helicity distributions,
although there are still a number of uncertainties concerning the flavor and helicity
structure of the sea quarks and also of the valence quarks at large Bjorken
variable $x \to 1$. High precision experiments on nucleons could in principle
explore these regions, and therefore eliminate these uncertainties. However,
it is also meaningful if we can find a new domain where the
same physics concerning the quark distribitions of the nucleon can manifest
itself in a way that is easy and clean to be detected and studied.
Indeed, the quark structure of the $\Lambda$ hyperon  
has become a new
area to study the quark structure of hadrons, and there has been continuous  
progress in this direction recently
[3-23]. There are also proposals to study the quark structure of
$\Sigma$ and $\Xi$ hyperons \cite{MSSY7,MSY4,Cao00}. Therefore it
is timely   
to study the quark distributions of other baryons, in
addition to those of the nucleon.

In principle, the precise knowledge of quark distributions of the  
other baryons should be measured from experiments, and then
compared with theoretical predictions. There have been a number of
predictions on the quark distributions of $\Lambda$ via
fragmentation processes \cite{MSY2,MSY3,MSSY5,MSSY6,MSSY7}, and on
quark distributions of $\Sigma$ and $\Xi$ through Drell-Yan
processes \cite{MSY4}. All of these predictions suffer, to  
some extent,   
from theoretical uncertainties. In fact, the quark
distributions of the octet baryons should be related to each other
from theoretical considerations, but the precise form of this  
relation is not known.
The purpose of this paper is to study the quark distributions of the  
other octet baryons by connecting them with the quark
distributions of the nucleon using SU(3) symmetry relations. For
the quark distributions of the nucleon we do not use theoretical
calculations, instead, we use available parametrizations. In this
way we are able to make realistic predictions concerning the
bulk features of physical quantities related to the quark
distributions of the other baryons. If these 
theoretical predictions disagree with future experimental measurements, then
we will be able to modify our approach in two possible ways. First, to  
change the relations used for connecting the quark distributions
between different baryons.
For example, we can consider SU(3) breaking effects between the
quark distributions of different baryons, and we should be able to
study and extract these SU(3) breaking effects with more
experimental data of quark distributions for different baryons.
Second, to use the experimental results of the quark  
distributions of other baryons to constrain the uncertainties of
quark distributions of the nucleon. For example, the discrepancies
between predictions and experiments may come from improper
assumptions in the quark distributions of the nucleon, and the
effects of these assumptions lead to incorrect behaviors  
of the quark distributions of the other octet baryons. Therefore we
can use the quark distributions of the other baryons as a laboratory  
to test the quark distributions of the nucleon.

In Section 2 we will use SU(3) symmetry to connect the  
quark distributions of the other octet baryons with those of the
nucleon. Using two available parametrizations of unpolarized and
polarized nucleon quark distributions, we find a new scenario for
the quark flavor and spin structure of the $\Lambda$, different
from two models already known:  a perturbative QCD based analysis and a  
quark spectator diquark model. In Section 3 we will make predictions  
of various quark to octet baryon fragmentations in several
processes, by using an approximate relation to connect the quark
distributions with fragmentation functions. It will be shown that
the predicted $\Lambda$ polarizations in several processes are
consistent with available data at large fragmentation momentum
fraction $z$, and predictions for other octet baryon polarizations
are also made. In Section 4, we present our predictions for the cross  
sections in Drell-Yan processes induced by $\Sigma^{\pm}$ and
$\Xi^{-}$ beams. The quark distributions of the $\Sigma^{\pm}$ and
$\Xi^{-}$ hyperons are directly used in the calculations,
therefore these processes are most suitable to explore  the sea
quark content of the octet baryons and the SU(3) relations between
octet baryons. Section 5 is devoted to some discussions and summary.

\section{Quark Distributions of Octet Baryons}

There have been a number of parametrizations of the unpolarized
\cite{CTEQ5,GRVupol,MRSTupol} and polarized
\cite{GRSVpol,GSpol,AACpol,GGGpol,BBpol} quark distributions of
the nucleon. In Section 3 we will study the baryon  
polarization in quark fragmentation, and for this purpose we  
need both the unpolarized and polarized quark distributions of the
baryons. We choose two sets of parametrizations for both the
unpolarized and polarized quark distributions of the nucleon
\cite{GRVupol,MRSTupol,GRSVpol,GSpol}, in order to  
illustrate our procedure for extending the quark distributions
from the nucleon to other octet baryons, by using SU(3)  
symmetry relations. Although the quark-antiquark pairs in the baryon
may have non-trivial behaviors, such as quark-antiquark asymmetry
\cite{Bro96}, it is still common practice to assume  
quark-antiquark symmetry, in the available standard
parametrizations of quark distributions. Therefore we will assume
the following symmetry relations for the sea quark distributions
in the baryon $B$
\begin{eqnarray}
q^B_s(x)=\bar{q}^{B}(x), ~~~ \Delta q^B_s(x)=\Delta \bar{q}^{B}(x),
\end{eqnarray}
where $q^B_{s}(x)$ means the sea quark distribution of  
$B$ and $\bar{q}^{B}(x)$ means the antiquark distribution of the
baryon sea. The valence quark distribution of the baryon is then
defined as
\begin{eqnarray}
q^B_v(x)=q^{B}(x)-\bar{q}^{B}(x),
~~~ \Delta q^B_v(x)=\Delta q^{B}(x)-\Delta \bar{q}^{B}(x),
\end{eqnarray}
where $q^{B}(x)$ means the quark distributions of both valence  
and sea quarks with flavor $q$ in $B$. In the  
parametrizations of quark distributions of the nucleon, only the
relations $q_v=q_v^p$ and $\bar{q}=\bar{q}^p$ for unpolarized, and
$\Delta q_v=\Delta q_v^p$ and $\Delta \bar{q}=\Delta \bar{q}^p$
for polarized ones, are needed for a whole set of quark
distributions.

The SU(3) symmetry relations between
the quark distributions of $p$, $n$, $\Sigma^{\pm}$,
$\Xi^-$, and $\Xi^0$ can be simply obtained by permutations  
between $u \to d \to s$
and $\bar{u} \to \bar{d} \to \bar{s}$
\begin{equation}
\begin{array}{lllc}
u^p=d^n=u^{\Sigma^+}=d^{\Sigma^-}=s^{\Xi^-}=s^{\Xi^0}=u;\\
d^p=u^n=s^{\Sigma^+}=s^{\Sigma^-}=d^{\Xi^-}=u^{\Xi^0}=d;\\
s^p=s^n=d^{\Sigma^+}=u^{\Sigma^-}=u^{\Xi^-}=d^{\Xi^0}=s;\\
\bar{u}^p=\bar{d}^n=\bar{u}^{\Sigma^+}=\bar{d}^{\Sigma^-}
=\bar{s}^{\Xi^-}=\bar{s}^{\Xi^0}=\bar{u};\\
\bar{d}^p=\bar{u}^n=\bar{s}^{\Sigma^+}=\bar{s}^{\Sigma^-}
=\bar{d}^{\Xi^-}=\bar{u}^{\Xi^0}=\bar{d};\\
\bar{s}^p=\bar{s}^n=\bar{d}^{\Sigma^+}=\bar{u}^{\Sigma^-}
=\bar{u}^{\Xi^-}=\bar{d}^{\Xi^0}=\bar{s}.
\end{array}
\end{equation}
For $\Sigma^0$ we notice the relation  
\begin{equation}
\Sigma^0=\frac{1}{2}(\Sigma^+ + \Sigma^-),
\end{equation}
therefore we get
\begin{equation}
\begin{array}{lllc}
u^{\Sigma^0}=
\frac{1}{2}( u^{\Sigma^+}+u^{\Sigma^-}) =\frac{1}{2}( u+s);\\
d^{\Sigma^0}=
\frac{1}{2}( d^{\Sigma^+}+d^{\Sigma^-}) =\frac{1}{2}( u+s);\\
s^{\Sigma^0}=\frac{1}{2} (s^{\Sigma^+}+s^{\Sigma^-})=d;\\
\bar{u}^{\Sigma^0}=\frac{1}{2}( \bar{u}^{\Sigma^+}+\bar{u}^{\Sigma^-})=\frac{1}{2} (\bar{u}+\bar{s});\\
\bar{d}^{\Sigma^0}=\frac{1}{2}( \bar{d}^{\Sigma^+}+\bar{d}^{\Sigma^-})=\frac{1}{2} (\bar{u}+\bar{s});\\
\bar{s}^{\Sigma^0}=\frac{1}{2}( \bar{s}^{\Sigma^+}+\bar{s}^{\Sigma^-})=\bar{d}.
\end{array}
\end{equation}
For $\Lambda$, the SU(3) relation connecting its valence  
quarks with those of the proton \cite{MSSY7} is
\begin{equation}
\begin{array}{lllc}
u_v^{\Lambda}=d_v^{\Lambda}=\frac{1}{6} u_v + \frac{4}{6} d_v;\\
s_v^{\Lambda}=\frac{2}{3} u_v - \frac{1}{3} d_v.
\end{array}
\end{equation}
We also know from the baryon-meson fluctuation picture of the
intrinsic sea quark-antiquark pairs \cite{Bro96}, that  
$\Lambda$ should have a similar sea structure as  
$\Sigma^0$, therefore we use
\begin{equation}
\begin{array}{lllc}
\bar{u}^{\Lambda}=\bar{u}^{\Sigma^0}=\frac{1}{2} (\bar{u}+\bar{s});\\
\bar{d}^{\Lambda}=\bar{d}^{\Sigma^0}=\frac{1}{2} (\bar{u}+\bar{s});\\
\bar{s}^{\Lambda}=\bar{s}^{\Sigma^0}=\bar{d}.
\end{array}
\end{equation}
Although there might be some model-dependence in the above
relations for the sea of $\Lambda$, such relations can be  
investigated and checked by looking at the quark distributions of
 $\Lambda$ at small $x$. Nevertheless, this will not   
affect much our predictions for the $\Lambda$ polarization  
in quark fragmentation, since these predictions are only  
 reliable at large $x$, a region depending mainly on  
the valence structure of the $\Lambda$ (see Section 3). Therefore we  
have all the necessary relations for extending the quark  
distributions of the nucleon to those of the other members of the  
octet baryons, and we list the above results in Table 1. The
formulae for the polarized quark distributions can be simply  
obtained by adding $\Delta$ in front of the unpolarized ones. We
notice that $s-\bar{s}=0$ in the standard quark distributions
with a quark-antiquark symmetric sea.  

\vspace{0.8cm}

\centerline{Table 1~~ The  quark distributions of the octet  
baryons from SU(3) symmetry}
\vspace{0.1cm}
\begin{scriptsize}
\begin{center}
\begin{tabular}{|c|c|c|c||c|c|c|}\hline
 Baryon & $u^B_v$ & $d^B_v$ & $s^B_v$ & $\bar{u}^B$ &
 $\bar{d}^B$ & $\bar{s}^B$  \\ \hline
p & $u_v$ & $d_v$ & $s-\bar{s}$ & $\bar{u}$ & $\bar{d}$ & $\bar{s}$  \\ \hline
n & $d_v$ & $u_v$ & $s-\bar{s}$ & $\bar{d}$ & $\bar{u}$ & $\bar{s}$   \\ \hline
$\Sigma^{+}$ & $u_v$ & $s-\bar{s}$ & $d_v$ & $\bar{u}$ & $\bar{s}$ & $\bar{d}$  \\ \hline
$\Sigma^{0}$ & $\frac{u_v+s-\bar{s}}{2}$ & $\frac{u_v+s-\bar{s}}{2}$ & $d_v$ &
$\frac{\bar{u}+\bar{s}}{2}$ &
$\frac{\bar{u}+\bar{s}}{2}$ & $\bar{d}$   \\ \hline
$\Sigma^{-}$ & $s-\bar{s}$ & $u_v$ & $d_v$ & $\bar{s}$ & $\bar{u}$ & $\bar{d}$
\\ \hline
$\Lambda^{0}$ & $\frac{u_v+4d_v}{6}$ & $\frac{u_v+4d_v}{6}$ & $\frac{2u_v-d_v}{3}$ &
 $\frac{\bar{u}+\bar{s}}{2}$ &
$\frac{\bar{u}+\bar{s}}{2}$ & $\bar{d}$    \\ \hline
$\Xi^{-}$ & $s-\bar{s}$ & $d_v$ & $u_v$ & $\bar{s}$ & $\bar{d}$ & $\bar{u}$  \\ \hline
$\Xi^{0}$ & $d_v$ & $s-\bar{s}$ & $u_v$
& $\bar{d}$ & $\bar{s}$ & $\bar{u}$  \\ \hline
\end{tabular}
\end{center}
\end{scriptsize}

\vspace{0.5cm}

In fact, the results in Table 1 can be applied to 
any set of parametrized nucleon quark distributions,
in order to get the quark distributions of other octet baryons. 
As an example of the application,  
we will choose two sets of parametrizations for both the  
unpolarized and polarized quark distributions of the nucleon.\\

The Set-1 unpolarized and polarized nucleon quark distributions are
taken from Ref.~\cite{GRVupol} 
(with ISET=2, NLO, $\bar{\mathrm MS}$-Scheme) and
Ref.~\cite{GRSVpol} (with ISET=1, NLO, $\bar{\mathrm MS}$-Scheme), 
respectively. The
Set-2 unpolarized and polarized nucleon quark distributions are
taken from Ref.~\cite{MRSTupol} (With Mod=1, Set=COR01, central
gluon and $\alpha_s$) and  Ref.~\cite{GSpol} (with Iflag=0, gluon set
A), respectively.\\

These quark distributions may not fully match at large $x$,  
because from the lack of data in this region, the  
behaviors for $x \to 1$ are not well determined, so that the relations  
between the unpolarized and polarized parametrizations are not  
well controlled.  
The unpolarized nucleon quark distributions are well constrained
by a large amount of data, so that different  
sets of quark distributions should be in close agreement.  
However the difference between different sets of  
polarized quark distributions can be larger, since at the $x \to 1$
end-point, the flavor structure of the polarized quark distributions is  
still not well known experimentally. The flavor and helicity
structure of the two sets of nucleon quark distributions is
reflected in the ratios $d(x)/u(x)$, $\Delta u(x)/u(x)$, and
$\Delta d(x)/d(x)$, shown in Fig.~\ref{mssy11f1}. We notice that
$d(x)/u(x) \to 0$ as $x \to 1$, in agreement  
with the quark-diquark model prediction \cite{Fey72,DQM,Ma96}.
However, $\Delta u(x)/u(x) < 0.8$ as $x \to 1$ in the two sets of
nucleon quark distributions, with some difference from both the
quark-spectator-diquark (quark-diquark) model \cite{Ma96} and the
perturbative QCD based analysis \cite{countingr,Bro95}, which both  
predict $\Delta u(x)/u(x) \to 1$ as $x \to 1$. We also  
notice that $\Delta d(x)/d(x)$ in the range $[-0.4,-0.2]$
as $x \to 1$ for the two sets  
and this is inconsistent with 
the perturbative QCD analysis, which predicts $\Delta d(x)/d(x)
\to 1$ as $x \to 1$ \cite{countingr,Bro95}, but roughly consistent with
the quark-diquark model, which
predicts $\Delta d(x)/d(x) \to -1/3$ as $x \to 1$ \cite{Ma96}. 
At large $x$, say $x \sim 0.9$,  
the behavior of Set-2 for $\Delta u(x)/u(x)$ and of both Set-1 and Set-2 of $\Delta  
d(x)/d(x)$ seem to be consistent with the quark-diquark model.
Nevertheless large uncertainties remain for  
the parametrizations of polarized quark distributions for $x \sim 1$, due to the absence
of data.

\begin{figure}
\begin{center}
\leavevmode {\epsfysize=3.5cm \epsffile{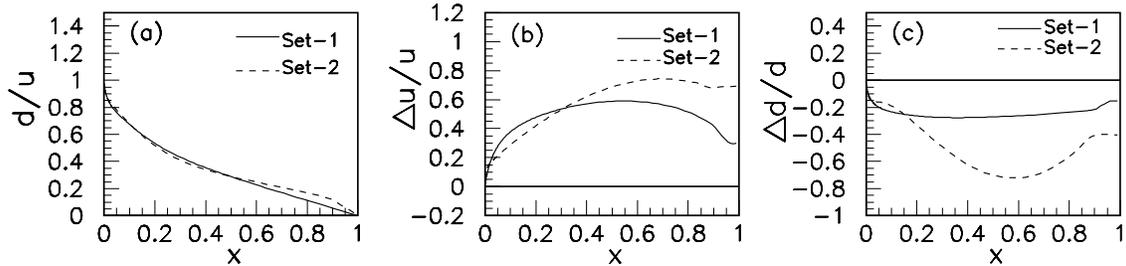}}
\end{center}
\caption[*]{\baselineskip 13pt The quark ratios of the
proton in the two sets of quark distributions: \\
(a) $d(x)/u(x)$;  (b)  $\Delta u(x)/u(x)$; (c) $\Delta d(x)/d(x)$. }
\label{mssy11f1}
\end{figure}

It has been suggested that the $\Lambda$ quark distributions at  
large $x$ are sensitive to different predictions of the  
flavor and helicity structure of the nucleon quark distributions.
By using the SU(3) relations in Table 1, we can get the quark
distributions of the $\Lambda$ from those of the nucleon, and we
present the ratios $u^{\Lambda}(x)/s^{\Lambda}(x)$, $\Delta
s^{\Lambda}(x)/s^{\Lambda}(x)$, and $\Delta
u^{\Lambda}(x)/u^{\Lambda}(x)$ in Fig.~\ref{mssy11f2}. It is
interesting to remark that we have  
$u^{\Lambda}(x)/s^{\Lambda}(x)=0.25$ at $x \to 1$, and this is
different from both the quark-diquark model prediction of $0$ and
the perturbative QCD prediction of $0.5$ \cite{MSY2,MSY3,MSSY7}.
This feature can be easily understood since at large $x \to 1$
\begin{equation}
u^{\Lambda}(x)/s^{\Lambda}(x) \approx u_v^{\Lambda}(x)/s_v^{\Lambda}(x)
= \frac{u_v+4d_v}{4u_v-2d_v}=\frac{1+4 d_v/u_v}{4-2d_v/u_v}
=\left\{
\begin{array}{ll}
\frac{1}{4} ~~~ {\mathrm for}
~~~\frac{d_v}{u_v}=0; \\
\frac{1}{2} ~~~ {\mathrm for}
~~~\frac{d_v}{u_v}=\frac{1}{5}.
\end{array}
\right.
\end{equation}
The above result of $\left. u_v^{\Lambda}(x)/s_v^{\Lambda}(x)
\right|_{x=1}=1/4$ for $d_v/u_v=0$ comes from the
SU(3) symmetry relations. We know that the quark-diquark model
predicts $\left. u_v^{\Lambda}(x)/s_v^{\Lambda}(x)
\right|_{x=1}=0$, although it also predicts $\left.
d_v(x)/u_v(x)\right|_{x=1}=0$ \cite{MSY2,MSY3,MSSY7}. The
difference between the above two different scenarios is due to the
fact that the quark-diquark model breaks SU(3) symmetry, and  
this breaking comes from the use of different quark and diquark
masses. In the perturbative QCD analysis, the helicity aligned
distributions (+) is dominant over the helicity anti-aligned (-), and  
therefore SU(6) is broken to SU(3)$^+ \times$SU(3)$^-$, which is a
different symmetry from the usually flavor SU(3) that we are
considering here. Therefore we arrive at the conclusion that the
flavor structure of the $\Lambda$ hyperon at large $x$ is also
sensitive to the SU(3) symmetry between different baryons, and it
is therefore an important region to test different
conjections and theories. The behavior of $\Delta  
u^{\Lambda}(x)/u^{\Lambda}(x)$ at large $x$ is also interesting,
since the SU(3) symmetry relations predict also positively  
polarized $u$ and $d$ quark distributions at large $x$, as can be
seen from Fig.~\ref{mssy11f2}~(b), and this coincides with the
predictions of both the quark-diquark model and the perturbative
QCD analysis. From the recent progress on $\Lambda$ production in
various quark fragmentation processes, we know that the prediction
of positively polarized $u$ and $d$ quarks inside the $\Lambda$ is
supported by all of the available data
\cite{MSY2,MSY3,MSSY5,MSSY6,MSSY7}. In the next section, 
we will present our predictions for baryon polarizations in  
various quark fragmentation processes using the above SU(3)
relations for the quark distributions of baryons.

\begin{figure}
\begin{center}
\leavevmode {\epsfysize=3.5cm \epsffile{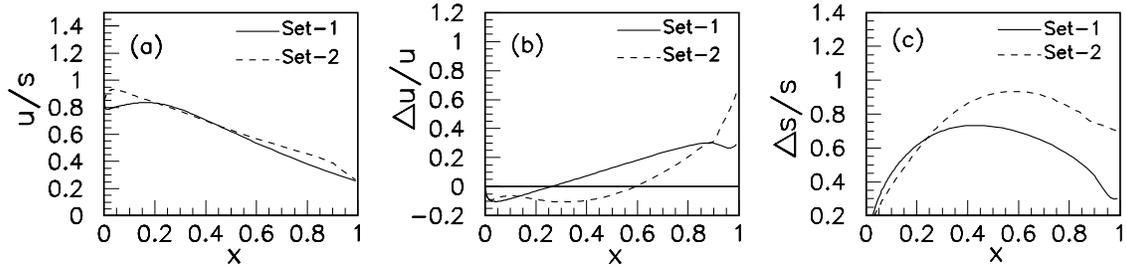}}
\end{center}
\caption[*]{\baselineskip 13pt The quark ratios of the
$\Lambda$ in the two sets of quark distributions: \\
(a) $u^{\Lambda}(x)/s^{\Lambda}(x)$;  (b)  $\Delta u^{\Lambda}(x)/u^{\Lambda}(x)$;
(c) $\Delta s^{\Lambda}(x)/s^{\Lambda}(x)$. }
\label{mssy11f2}
\end{figure}

\section{Baryon Polarizations in Quark Fragmentations}

Although the quark structure of the $\Lambda$ has some significant
features which can be used to distinguish between different
predictions, it is difficult to measure the quark distributions of
the $\Lambda$ directly, since the $\Lambda$ is a charge
neutral-particle which cannot be accelerated as incident beam and
its short lifetime makes it also difficult to be used as a target. However,
there have been attempts to connect the quark distributions with
the quark fragmentation functions, so that one can use hadron
productions from quark fragmentations to check the quark stucture
of hadrons. The connection is the so called Gribov-Lipatov (GL)
relation \cite{GLR}
\begin{equation}
D_q^h(z) \sim z q_h(z),
\label{GLR}
\end{equation}
where $D_q^h(z)$ is the
fragmentation function for a quark $q$ splitting into a hadron $h$
with longitudinal momentum fraction $z$, and $q_h(z)$ is the quark
distribution of finding the quark $q$
carrying a momentum fraction $x=z$ inside the hadron $h$. The GL relation should
be considered as an approximate relation near $z \to 1$ at an input energy scale $Q^2_0$  
~\cite{BRV00,Bar00}.
It is interesting to note that such a relation provided
successful descriptions of the available $\Lambda$ polarization 
data in several processes
\cite{MSY2,MSY3,MSSY5,MSSY6,MSSY7},
based on quark distributions of the $\Lambda$ in the
quark diquark model and in the pQCD based counting rule
analysis.
Thus we still
use (\ref{GLR}) as an Ansatz to relate the quark fragmentation
functions to the corresponding quark
distributions. This may be understood as a phenomenological method to parametrize the quark
fragmentation functions, and then we can check and improve these fragmentation functions
by comparing the predictions with experimental observations.  
To reduce the uncertainties in the GL relation,
we will only predict the baryon polarizations,
rather than the absolute values of cross sections.

\subsection{Baryon Polarizations in $e^+e^-$ Annihilation}

In the standard model of electroweak interactions, the quarks and
antiquarks produced in unpolarized $e^+e^-$-annihilation near
the Z pole should be
polarized due to the parity-violating coupling of the fermions,
and this leads to the polarizations of the hadrons produced in
quark fragmentations. The hadron polarization in
$e^+e^-$-annihilation can be written as
\begin{equation}
P_{h}=-\frac{\sum\limits_{q} A_q [\Delta D_q^h(z)-\Delta D_{\bar
q}^h(z)]}{\sum\limits_{q} C_q [D_q^h(z)+D_{\bar q}^h(z)]},
\label{PL2}
\end{equation}
where $A_q$ and $C_q$ are determined by the  
standard model. Explicit expressions can be found in
Refs.~\cite{MSY2,MSSY7}. $D_q^h(z)$ and $\Delta D_q^h(z)$ are the
unpolarized and polarized fragmentation functions for the quark
with flavor $q$ splitting into hadron $h$. Using the GL relation,
we can then calculate the hadron polarizations for the octet
baryons with the two sets of octet baryon quark distributions
described in the last section. Uncertainties in the absolute
magnitude of fragmentation functions can be reduced, since the
hadron polarization (\ref{PL2}) only involves ratios between
different fragmentation functions. In Fig.~\ref{mssy11f3} we
present our predictions for the longitudinal hadron polarizations
of the octet baryons in $e^+e^-$-annihilation at two energies: LEP
I at the $Z$ resonance $\sqrt{s} \approx 91$~GeV and LEP II at
$\sqrt{s} \approx 200$~GeV. The available experimental data of the
$\Lambda$ are taken at the $Z$ resonance
\cite{ALEPH96,DELPHI95,OPAL97}, and from the figure we find that
the calculated results are compatible with the data. This  
supports the prediction of positively polarized $u$ and $d$ quarks
inside the $\Lambda$ at large $x$, similar to the results in
Refs.~\cite{MSY2,MSSY7}. More experimental data for the
octet baryon polarizations in $e^+e^-$ annihilation are necessary 
in order to check different predictions.

\begin{figure}
\begin{center}
\leavevmode {\epsfysize=15cm \epsffile{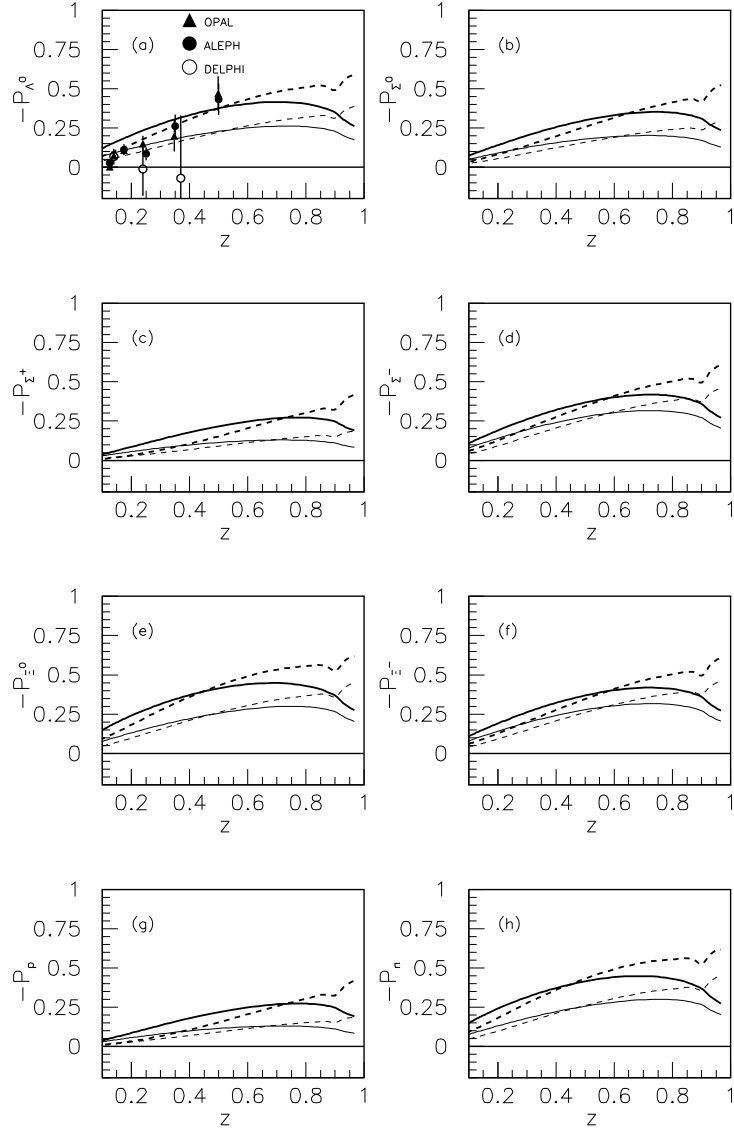}}
\end{center}
\caption[*]{\baselineskip 13pt
The predictions of the longitudinal
hadron polarizations for the octet baryons in
$e^+e^-$-annihilation at two energies: LEP I at $Z$ resonance
$\sqrt{s} \approx 91$~GeV (thick curves) and LEP II at $\sqrt{s}
\approx 200$~GeV (thin curves), with the input fragmentation functions
from the SU(3) symmetry of quark distributions for the octet baryons by the
the Gribov-Lipatov relation \cite{GLR}.
The solid curves and the dashed curves correspond to the
Set-1 and Set-2 quark
distributions of the nucleon, respectively.
The experimental data are taken from Refs.~\cite{ALEPH96,DELPHI95,OPAL97} } \label{mssy11f3}
\end{figure}

\subsection{Baryon Polarizations in Charged Lepton DIS Process}

In deep inelastic scattering of a longitudinally polarized
charged lepton on an unpolarized nucleon target, the scattered
quark will be polarized and its spin will be transferred to the
baryon produced in the fragmentation of this quark. The longitudinal spin transfer  
to the outgoing hadron $h$ is given in the quark-parton model by  
\begin{equation}
A^{h}(x,z)= \frac{\sum\limits_{q} e_q^2 [q^N(x,Q^2) \Delta
D_q^h(z,Q^2) + ( q \rightarrow \bar q)]}{\sum\limits_{q} e_q^2
[q^N (x,Q^2) D^h_q(z,Q^2) + ( q \rightarrow \bar q)]}~,
\label{DL}
\end{equation}
where a detailed description of the quantities in the above
formula can be found in Refs.~\cite{Jaf96,MSY2,MSSY5,MSSY6}. We
can also calculate the spin transfers for the octet baryons with
the two sets of quark distributions as input, and we present our
predictions in Figs.~\ref{mssy11f4} and \ref{mssy11f5}. The final
detected hadron could be either a baryon or an anti-baryon, and a
combination of both data provide information of
quark and antiquark to hadron fragmentations. Therefore we present
our predictions of the spin transfers for both baryons and
antibaryons respectively. There is some preliminary data by the
HERMES Collaboration \cite{HERMES} on $\Lambda$ production, and by
E665 Collaboration \cite{E665} on $\Lambda$ and $\bar{\Lambda}$
productions. The E665 data are too rough, and mainly focused on
the small $z$ region where the GL relation is not expected to
work well. Our calculated results based on the SU(3) relations of quark
distributions are consistent with the HERMES point, and this  
supports the prediction of positively polarized $u$ and $d$ quarks
inside the $\Lambda$. However, the $\Xi^0$ (and $\bar{\Xi}^0$)
polarizations are predicted to be negative at large $z$, and this
differs from either the quark-diquark model and the perturbative
QCD analysis, as can be seen by compare Figs.~\ref{mssy11f4}(e)  
and \ref{mssy11f5}(e) with Figs.~2(e) and 3(e) in  
Ref.~\cite{MSSY7}. There are several
reasons for negative $\Xi^0$ (and $\bar{\Xi}^0$) polarizations.  
First, $u \to \Xi^0$ is a dominant fragmentation chain as  
a result of the dominant $u$ quarks of the target and the squared
charge factor of $4/9$. Second, the $u$ quarks are negatively  
polarized inside the $\Xi^0$ and also the ratio
$u^{\Xi^-}(x)/s^{\Xi^-}(x)=1/4$ is not negligible when $x \to  
1$. Thus $\Xi^0$ and $\bar{\Xi}^0$ polarizations in polarized DIS
process are sensitive physical quantities that can distinguish
between different scenarios concerning the flavor and helicity
structure of the octet baryons at large $x$ (For similar and
detailed discussions, see Refs.~\cite{MSSY5,MSSY6,MSSY7}).

\begin{figure}
\begin{center}
\leavevmode {\epsfysize=15cm \epsffile{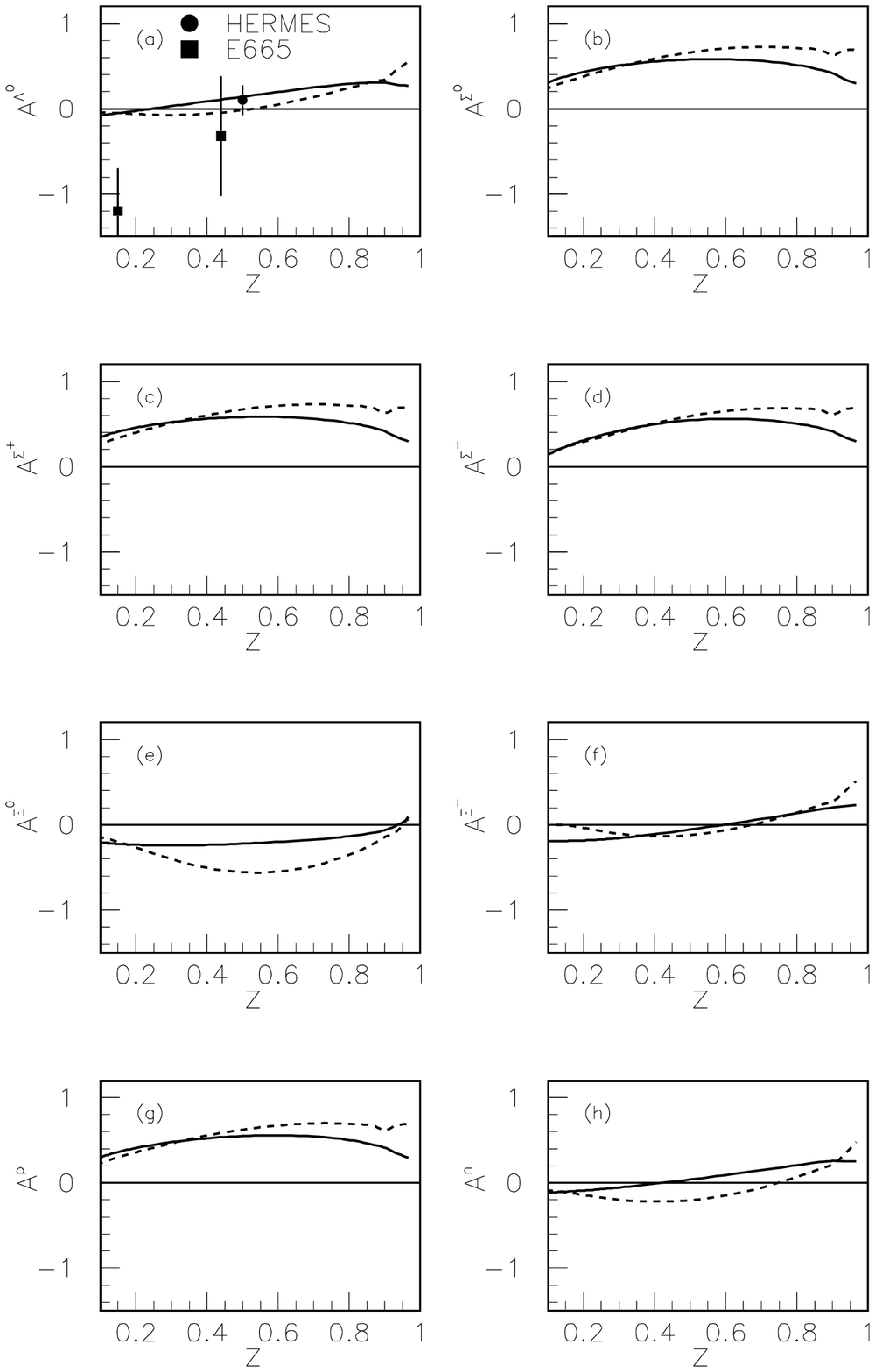}}
\end{center}
\caption[*]{\baselineskip 13pt The predictions of the
$z$-dependence for the {\it hadron} spin transfers
of the octet baryons in polarized charged
lepton DIS process on the proton target. The input fragmentation functions
are from the SU(3) symmetry of quark distributions for the octet baryons by the
the Gribov-Lipatov relation \cite{GLR}.
The solid curves and the dashed curves correspond to
the predictions using the
Set-1 and Set-2 quark
distributions of the nucleon. }\label{mssy11f4}
\end{figure}

\begin{figure}
\begin{center}
\leavevmode {\epsfysize=15cm \epsffile{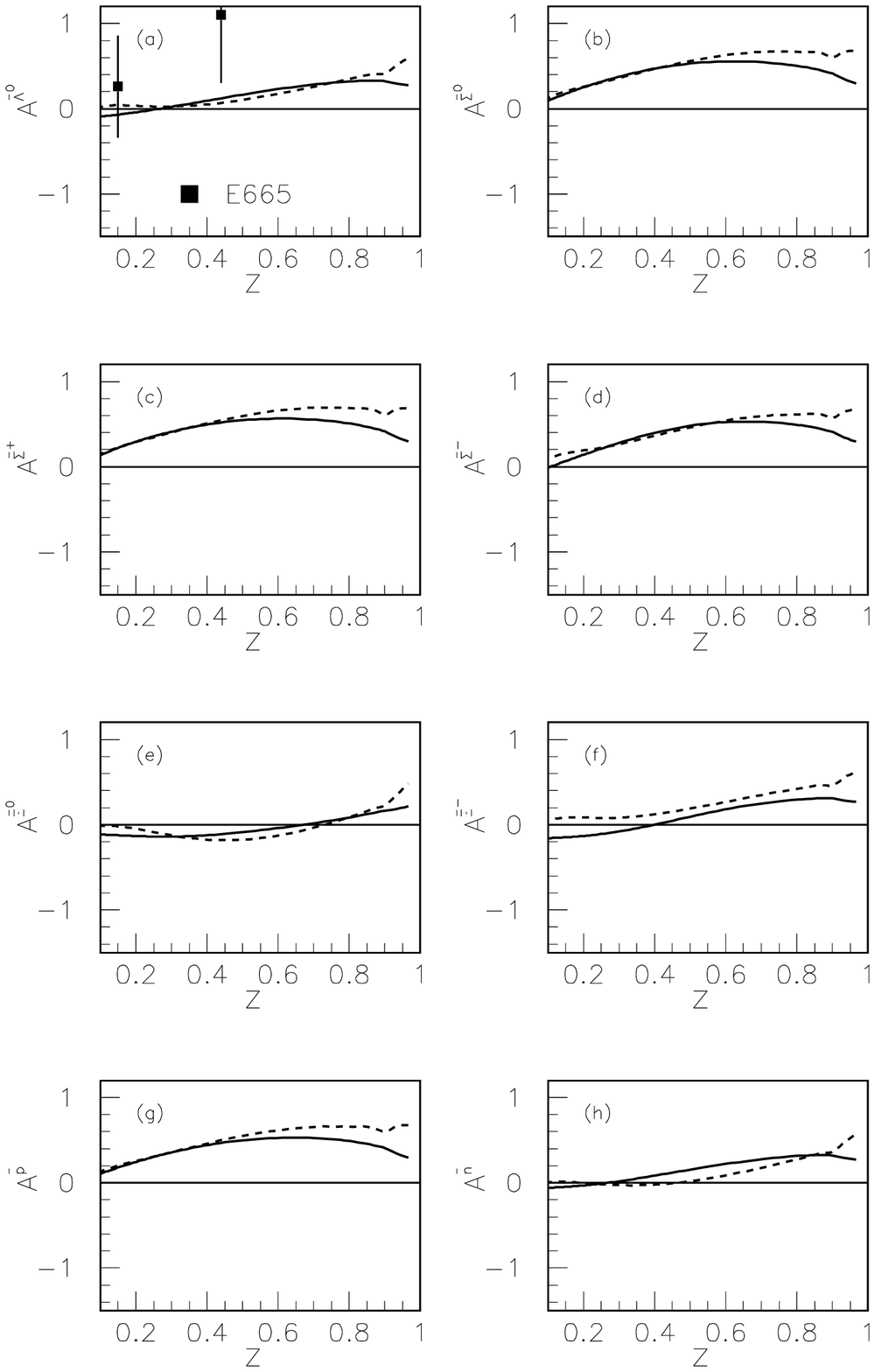}}
\end{center}
\caption[*]{\baselineskip 13pt The predictions of $z$-dependence
for the {\it anti-hadron} spin transfers of the
octet antibaryons in polarized charged lepton DIS
process on the proton target. The others are the same as
Fig.~\ref{mssy11f4}} \label{mssy11f5}
\end{figure}

\subsection{Baryon Polarizations in Neutrino DIS Process}

As has been pointed out in Refs.~\cite{Ma99,MSSY5}, the DIS scattering
of a neutrino beam on a hadronic target provides a source of
polarized quarks with specific flavor structure, and this makes it
an ideal system to study the flavor dependence of quark to hadron
fragmentation functions, especially in the polarized case. For the
production of any hadron $h$ from neutrino and antineutrino DIS
processes, the longitudinal polarization of $h$ in its momentum
direction, for $h$ in the current fragmentation region, can be
expressed as \cite{Ma99,MSSY5},
\begin{equation}
P_\nu^h(x,y,z)=-\frac{[d(x)+\varpi s(x)] \Delta D _u^h (z) -( 1-y)
^2 \bar{u} (x) [\Delta D _{\bar{d}}^h (z)+\varpi \Delta
D_{\bar{s}}^h(z)]} {[d(x)+\varpi s(x)] D_u ^h (z) + (1-y)^2
\bar{u} (x) [D _{\bar{d}}^h (z)+\varpi D_{\bar{s}}^h(z)]}~,
\end{equation}

\begin{equation}
P_{\bar{\nu}}^h (x,y,z)=-\frac{( 1-y) ^2 u (x) [\Delta D _d^h
(z)+\varpi \Delta D _s^h (z)]-[\bar{d}(x)+\varpi \bar{s}(x)]
\Delta D _{\bar{u}}^h (z)}{(1-y)^2 u (x) [D _d^h (z)+\varpi D _s^h
(z)]+[\bar{d}(x)+\varpi \bar{s}(x)] D_{\bar{u}} ^h (z)}~,
\end{equation}
where the terms with the factor $\varpi=\sin^2 \theta_c/\cos^2
\theta_c$ ($\theta_c$ is the Cabibbo angle) represent Cabibbo
suppressed contributions. The beam can be either neutrino or
antineutrino, and the produced hadron can be either baryon or
antibaryon. Therefore we have four combinations of different beams
and fragmented baryons, and they can provide different
information on the flavor dependence of quark fragmentation
functions. We present in Figs.~\ref{mssy11f6}-\ref{mssy11f13} the
longitudinal polarizations in the four different combinations of
beams and produced hadrons for each member of the octet baryons.
For $\Lambda$ production in neutrino DIS process, there have been
preliminary results by the NOMAD collaboration
\cite{NOMAD,Nomad00} which seem to support a positively polarized
$u$ and $d$ quarks inside the $\Lambda$ \cite{Yang01}. We notice
that our prediction for the $\Lambda$ polarization is compatible
with the data \cite{Nomad00} at large $z$, although there is one
experimental point at rather low $z$ that 
does not seem to be consistent with our results, as
can be seen in Fig.~\ref{mssy11f6}. Further studies are needed,
both theoretically and experimentally, concerning the detailed
features of $\Lambda$ production in neutrino DIS processes. More
precise measurements of $\Lambda$ ($\bar{\Lambda}$) polarization
in neutrino and antineutrino DIS processes seem hard to perform
before more intense neutrino beams or a neutrino factory
\cite{nufac} become available. Another realistic possibility is the  
production of $\Lambda$ in charged current semi-inclusive DIS at
HERA. We need to mention that several processes are dominated 
by antiquark contributions: $\Sigma^-$ and $\Xi^-$ productions in
neutrino process, and $\bar{\Sigma}^+$ and $\bar{\Xi}^+$ productions
in antineutrino process. In Figs.~(\ref{mssy11f9}) and (\ref{mssy11f11})
we remove the predictions for these processes
at $x>0.6$, due to the reason that the two parametrizations 
have some unphysical points with $|\Delta \bar{q}/\bar{q}|>1$ in this region.

\begin{figure}
\begin{center}
\leavevmode {\epsfysize=7cm \epsffile{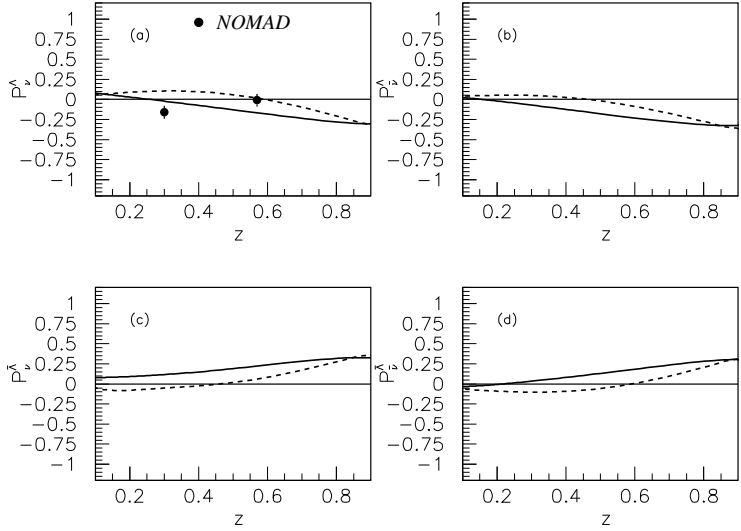}}
\end{center}
\caption[*]{\baselineskip 13pt The predictions of $z$-dependence
for the hadron and anti-hadron polarizations of $\Lambda$ in the
neutrino (antineutrino) DIS process. The input fragmentation functions
are from the SU(3) symmetry of quark distributions for the octet baryons by the
the Gribov-Lipatov relation.
The solid curves and the dashed curves correspond to the predictions
using the Set-1 and Set-2
quark distributions of the nucleon, for the proton target proton
$Q^2=4$~GeV$^2$ with the Bjorken variable $x$ integrated over
$0.02 \to 0.4$ and $y$ integrated over $0 \to 1$. }\label{mssy11f6}
\end{figure}

\begin{figure}
\begin{center}
\leavevmode {\epsfysize=7cm \epsffile{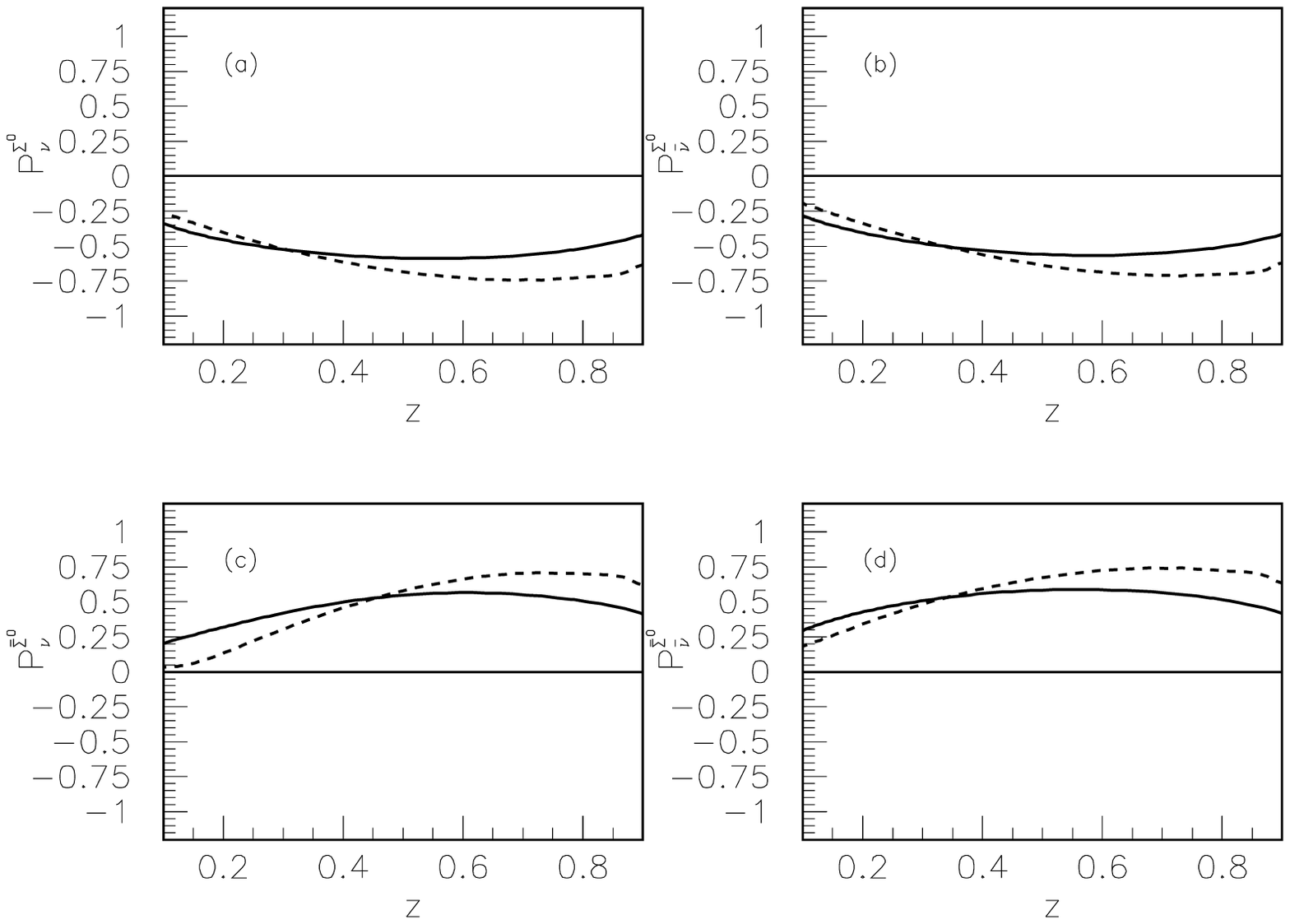}}
\end{center}
\caption[*]{\baselineskip 13pt The same as Fig.~\ref{mssy11f6},
but for predictions of $z$-dependence for the hadron and
anti-hadron polarizations of $\Sigma^0$ in the neutrino
(antineutrino) DIS process. }\label{mssy11f7}
\end{figure}

\begin{figure}
\begin{center}
\leavevmode {\epsfysize=7cm \epsffile{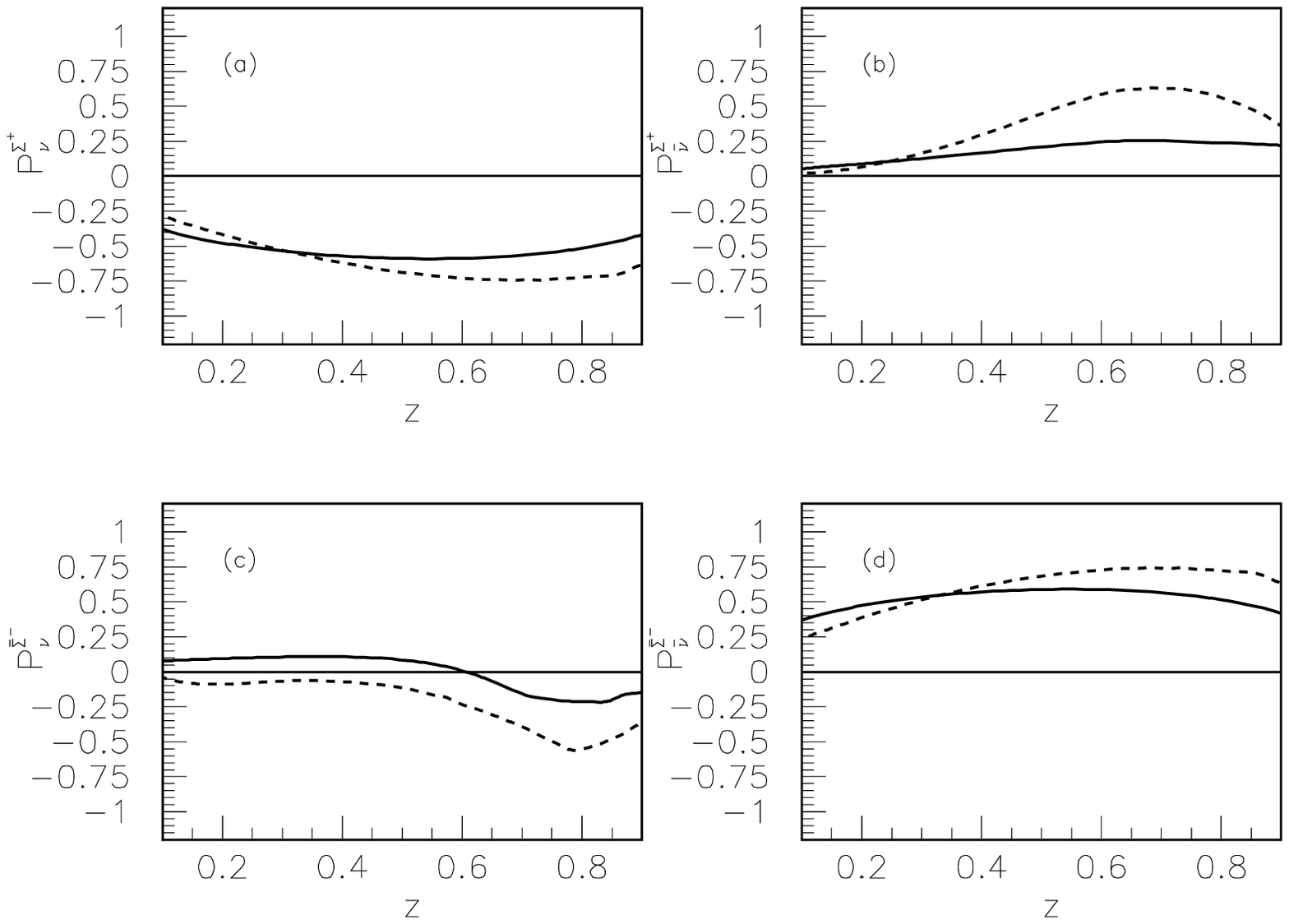}}
\end{center}
\caption[*]{\baselineskip 13pt The same as Fig.~\ref{mssy11f6},
but for predictions of $z$-dependence for the hadron and
anti-hadron polarizations of $\Sigma^+$ in the neutrino
(antineutrino) DIS process. }\label{mssy11f8}
\end{figure}

\begin{figure}
\begin{center}
\leavevmode {\epsfysize=7cm \epsffile{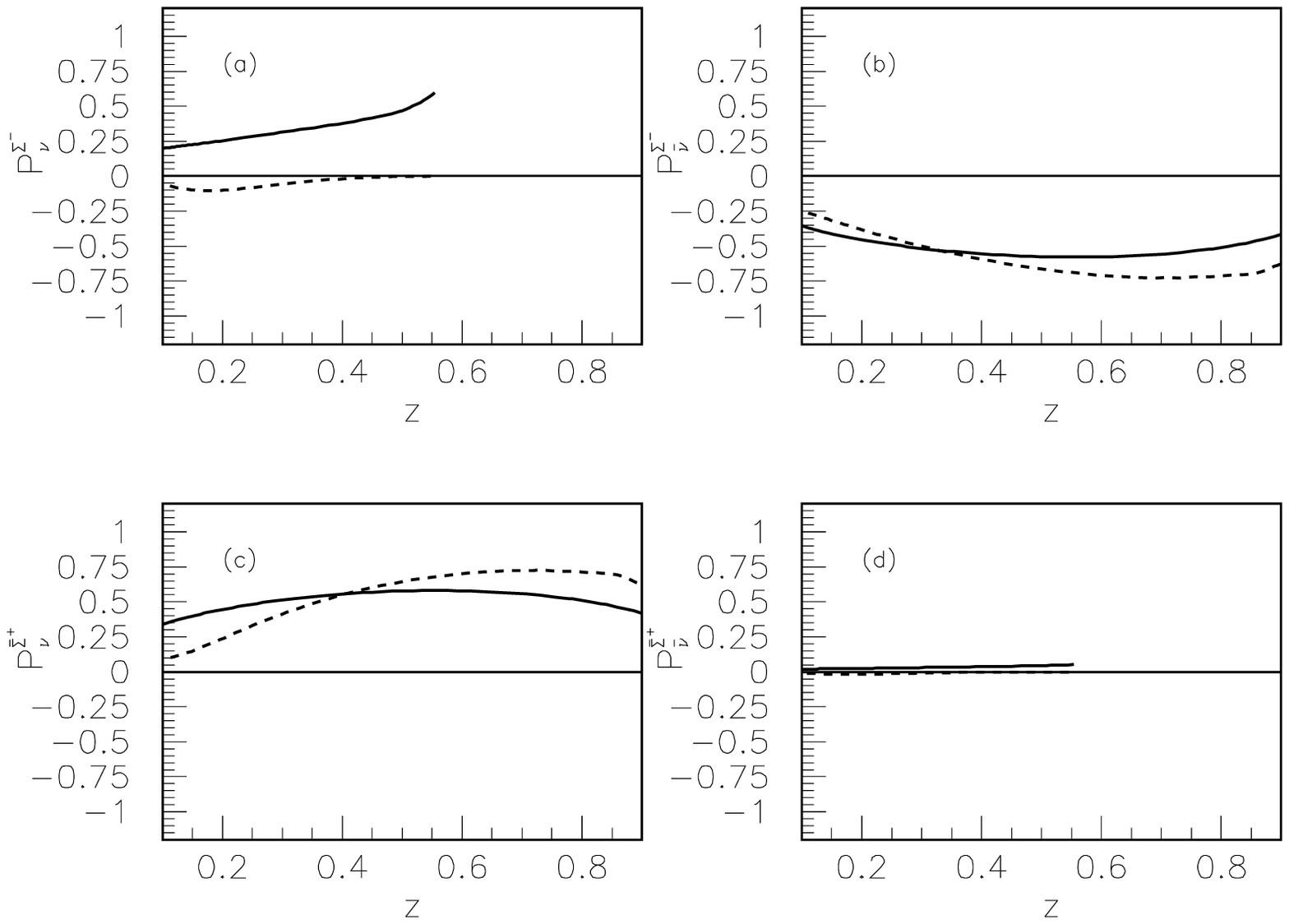}}
\end{center}
\caption[*]{\baselineskip 13pt The same as Fig.~\ref{mssy11f6}, but
for predictions of $z$-dependence for the hadron and anti-hadron
polarizations of $\Sigma^-$ in the neutrino (antineutrino) DIS
process. }\label{mssy11f9}
\end{figure}

\begin{figure}
\begin{center}
\leavevmode {\epsfysize=7cm \epsffile{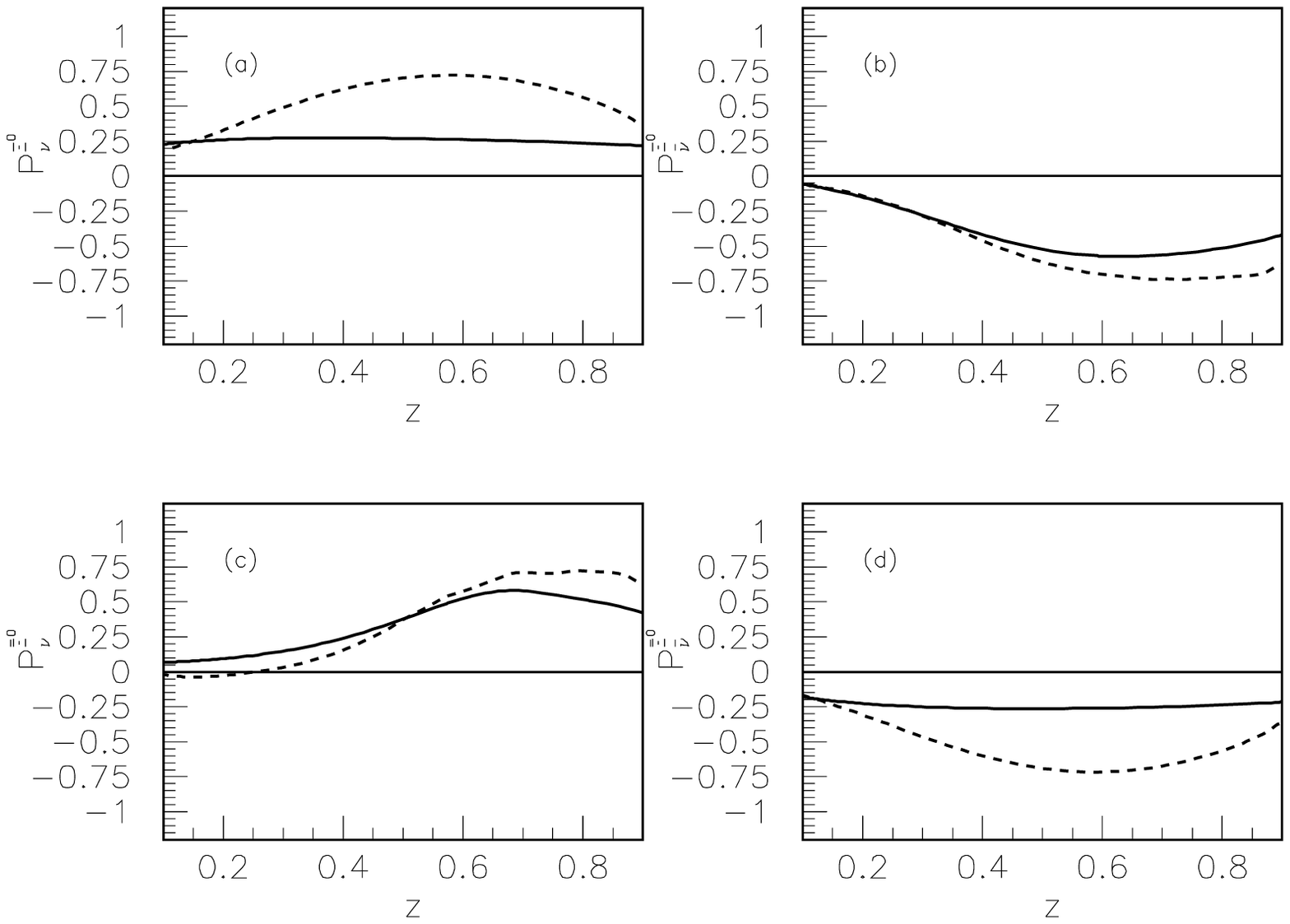}}
\end{center}
\caption[*]{\baselineskip 13pt The same as Fig.~\ref{mssy11f6},
but for predictions of $z$-dependence for the hadron and
anti-hadron polarizations of $\Xi^0$ in the neutrino
(antineutrino) DIS process. }\label{mssy11f10}
\end{figure}

\begin{figure}
\begin{center}
\leavevmode {\epsfysize=7cm \epsffile{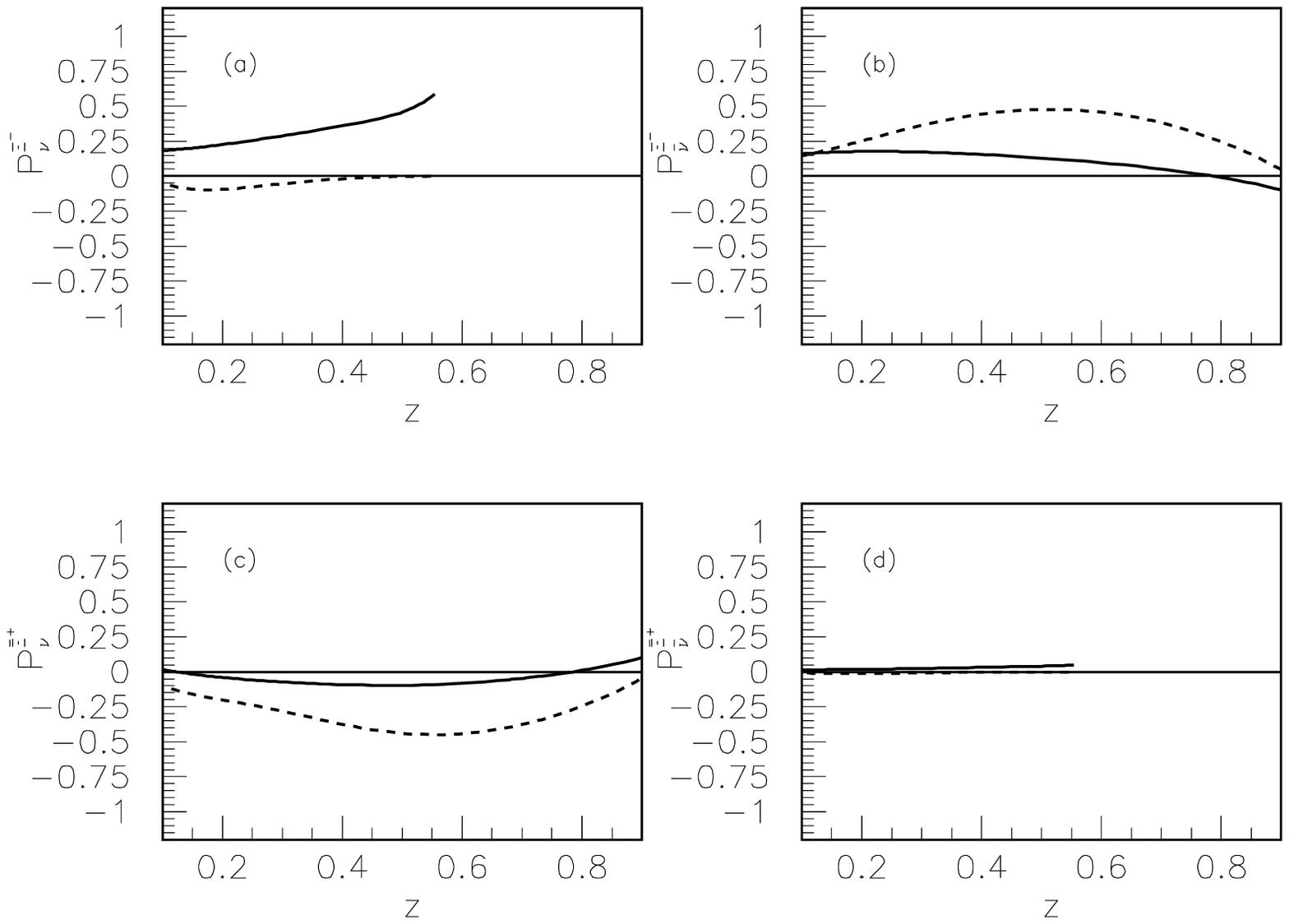}}
\end{center}
\caption[*]{\baselineskip 13pt The same as Fig.~\ref{mssy11f6},
but for predictions of $z$-dependence for the hadron and
anti-hadron polarizations of $\Xi^-$ in the neutrino
(antineutrino) DIS process. }\label{mssy11f11}
\end{figure}

\begin{figure}
\begin{center}
\leavevmode {\epsfysize=7cm \epsffile{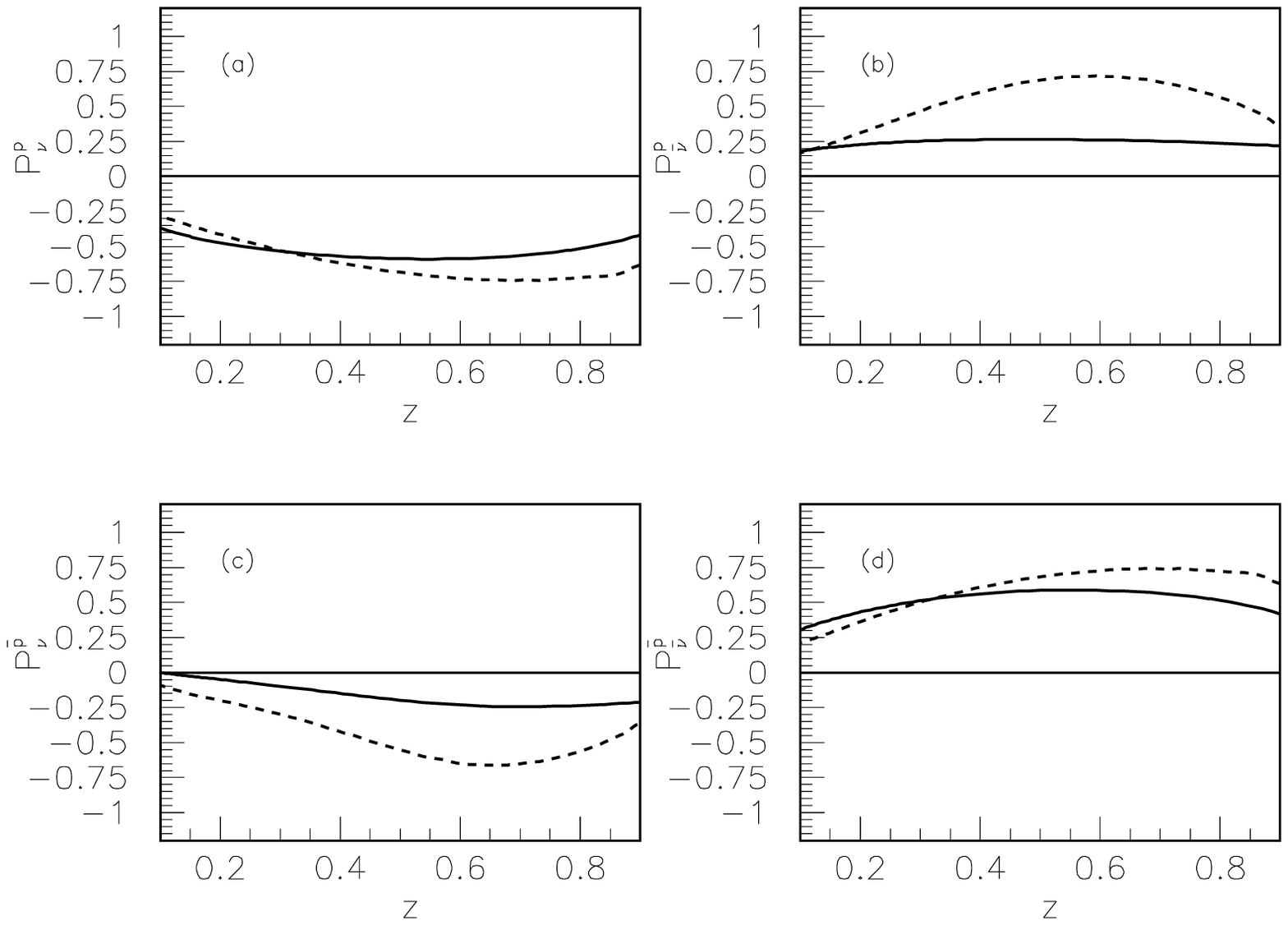}}
\end{center}
\caption[*]{\baselineskip 13pt The same as Fig.~\ref{mssy11f6},
but for predictions of $z$-dependence for the hadron and
anti-hadron polarizations of $p$ in the neutrino (antineutrino)
DIS process. }\label{mssy11f12}
\end{figure}

\begin{figure}
\begin{center}
\leavevmode {\epsfysize=7cm \epsffile{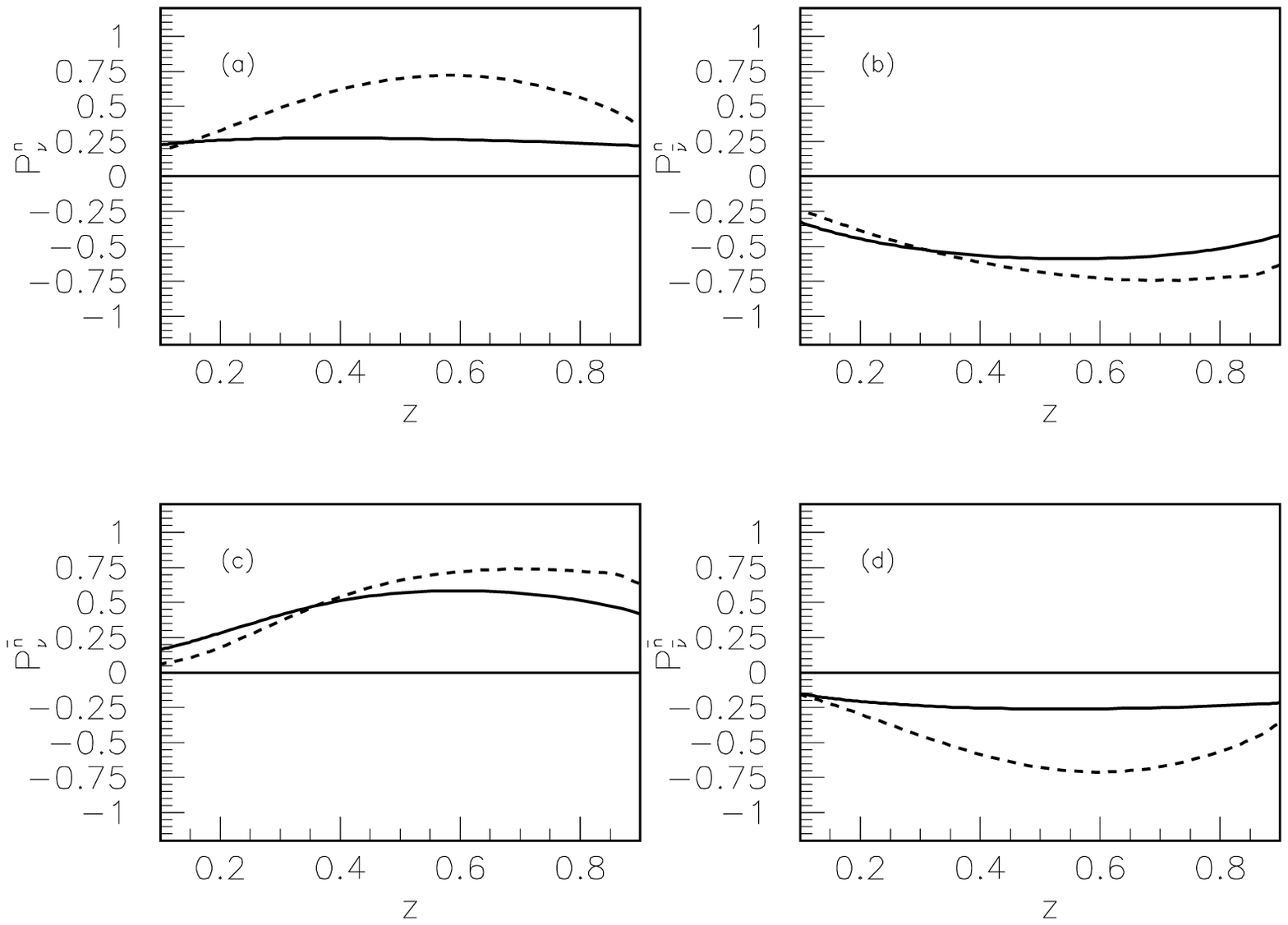}}
\end{center}
\caption[*]{\baselineskip 13pt The same as Fig.~\ref{mssy11f6},
but for predictions of $z$-dependence for the hadron and
anti-hadron polarizations of $n$ in the neutrino (antineutrino)
DIS process. }\label{mssy11f13}
\end{figure}

 From the above discussions of baryon polarizations in three different
processes, we arrive at the conclusion that all of the available
data of $\Lambda$ polarization support positively polarized
$u$ and $d$ quarks inside the $\Lambda$ at large $x$, and are
compatible with the theoretical predictions based on SU(3)
symmetric quark distributions. Of course, detailed features still
need further studies to confront theoretical understandings with
experimental observations (For some discussions, see
Refs.~\cite{MSY2,MSY3,MSSY5,MSSY6,MSSY7,MSSY8,MSY4}).

\section{Drell-Yan Process of $\Sigma^{\pm}$ Beams on Isoscalar
Target}

The above predictions of baryon polarizations in quark
fragmentations rely on the GL relation connecting the
fragmentation functions with the distribution functions, and such
a relation has not been firmly established, especially at small
$x$, although there have been some encouraging results to support
the use of this relation from the available $\Lambda$ polarization
data in several processes. Therefore in order to study the SU(3)
symmetry relations of the quark distributions between different
octet baryons, it is better to study the quark distributions
directly, instead of the fragmentation functions. It has been
proposed \cite{MSY4} that the Drell-Yan process of $\Sigma^{\pm}$
and $\Xi^-$ beams on the isoscalar target can be used to study the
quark distributions of $\Sigma^{\pm}$ and $\Xi^{-}$ hyperons. It
is the purpose of this section to perform such a study
by using the two sets of quark distributions for the octet baryons
described in Section 2.

For the Drell-Yan process
\begin{equation}
\Sigma N \to l^+ l^- X,
\end{equation}
the cross section can be written as
\begin{equation}
\sigma (\Sigma N)=\frac{8 \pi \alpha^2}{9 \sqrt{\tau}} K(x_1,x_2)
\sum\limits_f e_f^2 [q_f^{\Sigma}
(x_1)\bar{q}_f(x_2)+\bar{q}_f^{\Sigma}(x_1) q_f(x_2)],
\end{equation}
where $\sqrt{\tau}=M/\sqrt{s}$, $M$ is the mass of the dilepton
pair and $\sqrt s$ the total c.m. energy. The factor $K(x_1,x_2)$ 
is due to higher-order QCD  
corrections. For an isoscalar target with nucleon number $A$, we
obtain the cross sections  
\begin{eqnarray}
\sigma^{\pm}= \sigma(\Sigma^{\pm} A)=\frac{A}{2}
\left[\sigma(\Sigma^{\pm} p)+\sigma(\Sigma^{\pm} n) \right];
\label{S1}
\end{eqnarray}
from which we can obtain the ratio
\begin{equation}
T(x_1,x_2)=\frac{\sigma^+(x_1,x_2)}{\sigma^-(x_1,x_2)}.
\label{Tx12}
\end{equation}
Thus we can calculate this ratio with the two sets of quark
distributions of octet baryons. On the experimental side, it  
is convenient to use the variables $\tau$ and $y$, where $y$ is the rapidity  
of the dilepton pair, instead of the variables $x_1$ and $x_2$, 
through the relations  
\begin{eqnarray}
&&x_1=\rm{e}^y \sqrt{\tau}; \nonumber \\
&&x_2=\rm{e}^{-y}\sqrt{\tau}.
\end{eqnarray}
Thus we can express the quantity $T$ as a function of $\tau$ and
$y$.
We present our predictions for $T$ versus three different kinematic variables  
 in Figs.~\ref{mssy11f14}-\ref{mssy11f16}. The advantage of
these calculations is the fact that we know the complete quark  
distributions needed in $T$ from the two sets of quark
distributions, and therefore we do not need to make the
valence-dominance approximation in the calculations. 
We also present our calculation for $T$ with the
two sets of quark distributions by turning off 
the sea contributions in Eq.~(\ref{Tx12}), and find a big difference at
$x_1, x_2<0.7$ compared with the present calculation which
contains all quark contributions. This means that the sea quarks
play an important role in the Drell-Yan process and they cannot be
simply neglected in the numerical evaluation for $x_1, x_2<0.7$.

\begin{figure}[htb]
\begin{center}
\leavevmode {\epsfysize=4.5cm \epsffile{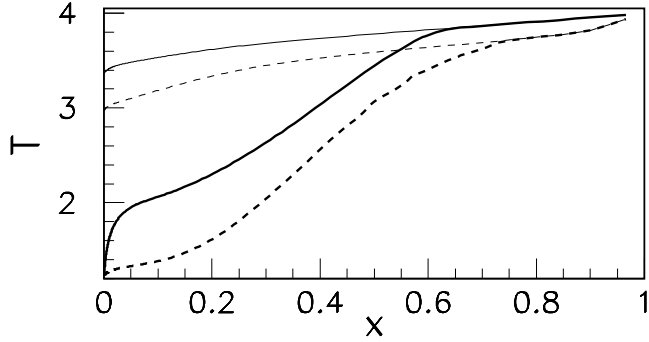}}
\end{center}
\caption[*]{\baselineskip 13pt The thick curves are $T(x_1,x_2)$
in the two sets of quark distributions at fixed $x_2=0.3$ as a
function of $x=x_1$, whereas the thin curves are the corresponding
$T(x_1,x_2)$ with only valence quark contributions.}\label{mssy11f14}
\end{figure}

\begin{figure}[htb]
\begin{center}
\leavevmode {\epsfysize=4.5cm \epsffile{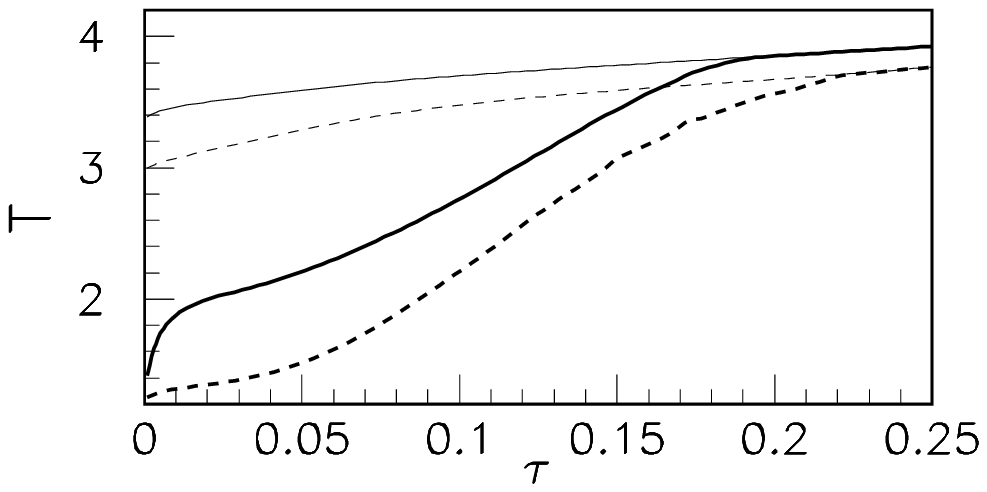}}
\end{center}
\caption[*]{\baselineskip 13pt The thick curves are $T(x_1,x_2)$
in the two sets of quark distributions at fixed $y=0$ as a
function of $\tau$, whereas the thin curves are the corresponding
$T(x_1,x_2)$ with only valence quark contributions.
 }\label{mssy11f15}
\end{figure}

\begin{figure}[htb]
\begin{center}
\leavevmode {\epsfysize=4.5cm \epsffile{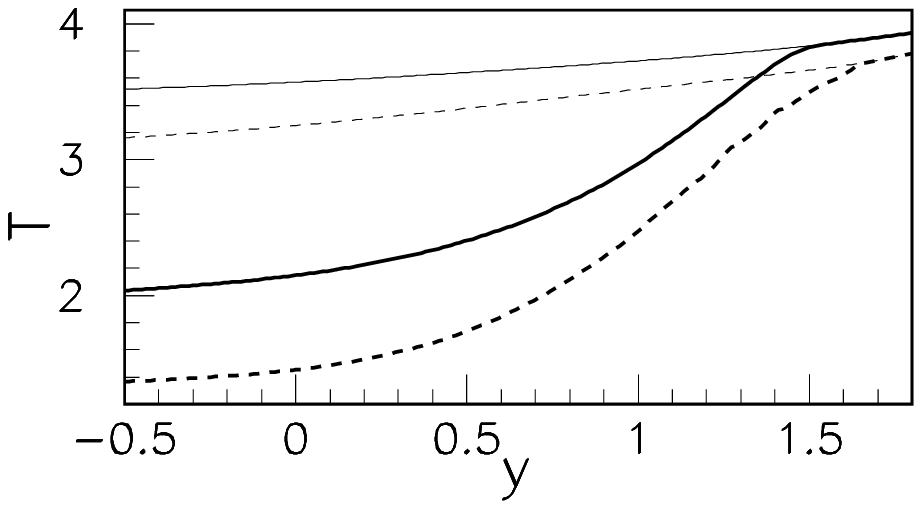}}
\end{center}
\caption[*]{\baselineskip 13pt The thick curves are $T(x_1,x_2)$
in the two sets of quark distributions at fixed $\tau=0.02$ as a
function of $y$, whereas the thin curves are the corresponding
$T(x_1,x_2)$ with only valence quark contributions.
 }\label{mssy11f16}
\end{figure}

Similar discussions and predictions can be also extended to the case where the
charged octet baryon $\Xi^-$ is used as the beam, i.e., for the Drell-Yan process
\begin{equation}
\Xi^{-} N \to l^+ l^- X.
\end{equation}
In this case $s$ is the dominant valence quark and $d$ is the less
dominant valence quark at large $x$ inside the $\Xi^-$, and this
can provide further information to check the flavor structure of
the two sets of quark distributions. We expect that the sea quarks also
play an important role in this case (For detailed formulae and  
discussions see Ref.~\cite{MSY4}).

\section{Discussions and Summary}

We showed in this paper that one can use SU(3) symmetry relations
between the octet baryons, in order to get a complete set of
unpolarized and polarized quark distributions of the octet baryons
from the known quark distributions of the nucleon.
Thus we have a new
domain to check the nucleon quark distribution parametrizations.

We found a new scenario of quark flavor and spin structure of the
$\Lambda$ in comparison with two already known models of a perturbative  
QCD based analysis and a quark diquark model. The $u$ and $d$
quarks inside the $\Lambda$ are predicted to be positively
polarized at large Bjorken variable $x$ in the new scenario.
Although the new scenario and the quark-diquark model have the
same ratio of $d(x)/u(x) \to 0$ for the proton at $x \to 1$, they
give very different predictions of the ratio
$u^{\Lambda}(x)/s^{\Lambda}(x)$ at $x \to 1$. The quark-diquark
model predicts $u^{\Lambda}(x)/s^{\Lambda}(x)=0$ whereas the SU(3)
symmetry predicts
$u^{\Lambda}(x)/s^{\Lambda}(x)=\frac{1}{4}$. The difference
between the two scenarios is due to the fact that the
quark-diquark model in fact breaks SU(3) symmetry, because it uses
different quark and diquark masses. Therefore we  
conclude that the flavor structure of the $\Lambda$ hyperon at
large $x$ is also sensitive to the SU(3) symmetry between
different baryons, and it is a region with rich physics to test
different models.

Using an approximate relation connecting the quark fragmentation
functions with the quark distributions, we predicted the hadron
polarizations of the octet baryons in several processes. The
prediction of the $\Lambda$ polarizations are compatible with the
available data at large fragmentation momentum fraction $z$, and
this supports the prediction of positively polarized $u$ and $d$
quarks inside the $\Lambda$ at large $x$. It is also shown that
the $\Xi^0$ and $\bar{\Xi}^0$ polarizations in polarized DIS
process are sensitive to different scenarios of the flavor and
helicity structure of the octet baryons at large $x$. We also  
presented predictions and discussions on Drell-Yan processes for
$\Sigma^{\pm}$ and $\Xi^{-}$ beams on isoscalar targets.

Our predictions can be used to check the SU(3)  
relations between the quark distributions of the octet baryons
and/or the quark distributions of the nucleon. There are still
uncertainties on the quark distributions of the nucleon concerning
its sea content and the flavor and helicity structure at large
Bjorken variable $x \to 1$. We have shown in this paper that one can  
reduce or eliminate these uncertanties by exploring the quark
structure of other members of the octet baryons. Thus systematic
studies of the quark distributions of the octet baryons will
introduce a new direction to confront and check our understandings
of the basic hadron structure by comparing theoretical predictions
with experimental observations.

{\bf Acknowledgments: }
This work is
partially supported by National Natural Science Foundation of
China under Grant Numbers 19975052, 19875024, and 10025523, by Fondecyt
(Chile) postdoctoral fellowship 3990048, by the cooperation
programmes Ecos-Conicyt C99E08 between France and
Chile, and by Fondecyt (Chile) grant 8000017.

\end{document}